\documentclass[traditabstract,twocolumn]{aa}

\usepackage{graphicx}
\usepackage{epsfig}
\usepackage{amssymb,amsmath}
\usepackage{natbib}
\bibpunct{(}{)}{;}{a}{}{,}
\usepackage{bibentry}
\usepackage{txfonts}
\usepackage{pslatex}
\usepackage{caption}
\usepackage{dcolumn}
\usepackage{bm}
\usepackage{subfig}

\newcommand\simlt{\lower.5ex\hbox{$\; \buildrel < \over \sim \;$}}
\begin{document}

\title{Investigation of dark matter-dark energy interaction cosmological model}

\author{ J. S. Wang\inst{1,2} and F. Y. Wang\inst{1,2}
}

\offprints{F. Y. Wang(fayinwang@nju.edu.cn)}

\institute{School of Astronomy and Space Science, Nanjing
University, Nanjing 210093, China; \and Key Laboratory of Modern
Astronomy and Astrophysics (Nanjing University), Ministry of
Education, Nanjing 210093, China}

\authorrunning{Wang \& Wang}
\titlerunning{Investigation of dark matter-dark energy interaction cosmological model}

\begin{abstract}
{In this paper, we test the dark matter-dark energy interacting
cosmological model with a dynamic equation of state
$w_{DE}(z)=w_{0}+w_{1}z/(1+z)$, using type Ia supernovae (SNe Ia),
Hubble parameter data, baryonic acoustic oscillation (BAO)
measurements, and the cosmic microwave background (CMB) observation.
This interacting cosmological model has not been studied before. The
best-fitted parameters with $1 \sigma$ uncertainties are
$\delta=-0.022 \pm 0.006$, $\Omega_{DM}^{0}=0.213 \pm 0.008$, $w_0
=-1.210 \pm 0.033$ and $w_1=0.872 \pm 0.072$ with $\chi^2_{min}/dof
= 0.990$. At the $1 \sigma$ confidence level, we find $\delta<0$,
which means that the energy transfer prefers from dark matter to
dark energy. We also find that the SNe Ia are in tension with the
combination of CMB, BAO and Hubble parameter data. The evolution of
$\rho_{DM}/\rho_{DE}$ indicates that this interacting model is a
good approach to solve the coincidence problem, because the
$\rho_{DE}$ decrease with scale factor $a$. The transition redshift
is $z_{tr}=0.63 \pm 0.07$ in this model.}
\end{abstract}


\maketitle

\section{Introduction}

Recent observations with increasing accuracy show that the universe
is undergoing an accelerating expansion, such as type Ia supernovae
(SNe Ia; Riess et al. 1998; Perlmutter et al. 1999; Suzuki et al.
2012), cosmic microwave background (CMB) from Wilkinson Microwave
Anisotropy Probe 9 years (WMAP9; Hinshaw et al. 2013) and Planck
\citep{Planck2013arXiv1303.5076P}, the baryonic acoustic oscillation
(BAO) from 6dF Galaxy Redshift Survey (6dFGRS; Beutler et al. 2011),
the Sloan Digital Sky Survey (SDSS; Eisenstein et al. 2005; Percival
et al. 2010; Anderson et al. 2012), WiggleZ \citep{Blake2012} and so
on. Planck results also confirm that the universe is
spatially flat, in other words, the curvature parameter $\Omega_K$ is
$-0.0000_{-0.0067}^{+0.0066}$ \citep{Planck2013arXiv1303.5076P} at
95\% confidence level. The main components of the universe are dark
matter (DM) and dark energy (DE). The special characteristic of DE
is negative pressure. The simplest candidate of DE is the
cosmological constant with equation of state (EoS)
$w=p_{\Lambda}/\rho_{\Lambda}=-1$. However, there are some problems
with the $\Lambda$CDM model. The most important one is coincidence
problem, which says why the DE density is comparable with the matter
density at present. Yet, the energy density of DE is non-dynamical
while matter density decreases with $a^{-3}$, where $a=1/(1+z)$ is
scale factor.

In order to solve the coincidence problem, many methods have been
proposed
\citep{Ratra1988,Caldwell2002,Armendariz-Picon2001,Feng2005}.
The interacting dark sectors models are possible solutions,
which means there is energy exchanges between DE and DM. So the energy density
ratio $ \rho_{DM} / \rho_{DE} $ can decrease slower than $a^{-3}$.
We consider that the energy exchanges through a interaction term
$Q$. The conservation of the total stress-energy tensor, and a
scalar-field model of dark energy is also assumed in this case
\begin{eqnarray}
(\dot{\rho}_{B} +\dot{\rho}_{DM} ) + 3H(\rho_{B} +\rho_{DM} ) =-Q , \label{dmevo}\\
\dot{\rho}_{DE}+ 3H\rho_{DE} (1+w_{DE}) = Q,\label{deevo}
\end{eqnarray}
where $\rho_{B}$ and $\rho_{DM}$ represent the energy density of
baryon and cold dark matter respectively, $\rho_{DE}$ is the energy
density of dark energy with a dynamic EoS $w_{DE}$, and
$H=\dot{a}/a$ is Hubble parameter. Many interacting theoretical
models have been studied \citep{Amendola2000,Farrar2004,Guo2005,
Szydlowski2006,Sadjadi2006,Del Campo2006,Wei2006,
Bertolami2007,Cai2010}. But the interaction term $Q$ is still poorly
known. Many phenomenological models have been put forward to solve
it, such as a simple phenomenological coupling form
$Q=C\delta(a)H\rho_{DM} $
\citep{Dalal2001,Amendola2007,Guo2007,Wei2010,Cao2011}, where $C$ is
constant. The EoS of dark energy is needed to solve the
Eq.\ref{deevo}. In this paper we will discuss a phenomenological
model with a dynamic EoS \citep{Chevallier2001,Linder2003},
\begin{equation}
w_{DE}(z)=w_{0}+w_{1}z/(1+z) .\label{EoS}
\end{equation}
Then we calculate the evolution of energy density of DM and DE.
The transition redshift is also constrained in this interacting phenomenological model
\citep{Abdel-Rahman2007}.

The structure of this paper is arranged as follows. In section 2, we
will analyze the model. In section 3, we constrain model parameters
using the observational data sets. In section 4, we present results.
The conclusions and discussions will be given in section 5.

\section{Interacting dark sector model}

The interaction is between the non-baryonic dark matter and the
quintessence field. The mass evolution of dark matter particles can
be written as $m=m(\Phi(a))$, and parameterize in a function of the
scale factor $\delta(a)$ \citep{Amendola2007,Majerotto2004,Rosenfeld2005},
\begin{equation}
m(a)=m_{0}e^{\int_{1}^{a}\delta(a')d\ln a'
},\label{eq:parametrization}\end{equation} where $m_{0}$ is the dark
matter mass today and $\delta(a)=d\ln m/d\ln a$ represents the rate
of change of the dark matter mass. We will set $\delta(a)$ as a constant in this paper \citep{Amendola2007}.

The evolutions of $\rho_{DM}$ and $\rho_{DE}$ can be expressed as
\begin{eqnarray}
\dot{\rho}_{DM}+3H\rho_{DM}-\delta H\rho_{DM}=0, \label{rhoDM}\\
\dot{\rho}_{DE}+3H\rho_{DE}(1+w_{DE})+\delta
H\rho_{DM}=0.\label{rhoDE}
\end{eqnarray}
The interacting term is $Q=-\delta H\rho_{DM}$. Then Eq.
(\ref{rhoDM}) can be solved in the assumption of a constant
interaction,
\begin{equation}
\rho_{DM}(a)=\rho_{DM}^{0}a^{-3+\delta},\label{DMSol}
\end{equation}
where $\rho_{DM}^{0}$ is the dark matter energy density today.
Substituting this solution into Eq. (\ref{rhoDE}), we obtain the
evolution of $\rho_{DE}$,
\begin{equation}
\frac{d\rho_{DE}}{da}+\frac{3}{a}\rho_{DE}(1+w_{DE})+\delta\rho_{DM}^{0}a^{-4+\delta}=0.\label{rhoDEa}
\end{equation}

Amendola et al. (2007) studied the interacting model with a EoS
$w_{DE}(z)=w_{0}+w_{1}z$. But that model is not compatible with CMB
data since it diverges at high redshift \citep{Chevallier2001}. We
consider an extended parameterization of EoS as Eq.(\ref{EoS}) to
avoid this problem \citep{Chevallier2001,Linder2003}. Then we obtain
solution of Eq. (\ref{rhoDEa}) as a function of redshift $z$,
\begin{equation}
\rho_{DE}(z)=\rho_{DE}^{NI}(z)\left[1+\Theta(z,w_{0},w_{1},\delta)\right] , \label{DESol}
\end{equation}
where
\begin{equation}
\rho_{DE}^{NI}(z)=\rho_{DE}^{0}e^{-3w_{1}z/(1+z)}(1+z)^{3(1+w_{0}+w_{1})}.
\end{equation}
It represents the evolution of dark energy density without
interaction for this parameterization. The $\Theta$ function is
\begin{eqnarray}
\Theta(z,w_{0},w_{1},\delta)=\delta\;e^{3w_{1}}(3w_{1})^{-3(w_{0}+w_{1})-\delta} \times\frac{\rho_{DM}^{0}}{\rho_{DE}^{0}}\;\nonumber
\\
~~~~~~~~~~~~~~~~~~~~~~\times\Gamma(3(w_{0}+w_{1})+\delta,3w_{1}/(1+z),3w_{1}),
\end{eqnarray}
where $\Gamma(a,x_{0},x_{1})$ is the generalized incomplete gamma
function
$\Gamma(a,x_{0},x_{1})=\int_{x_{0}}^{x_{1}}t^{a-1}e^{-t}dt.$

Then the Hubble parameter in this dark interaction model can be written as
\begin{eqnarray}
E(z,\Omega_{DM}^{0},w_{0},w_{1},\delta)=[\Omega_{DM}^{0}(1+z)^{3-\delta}+\Omega_{B}^{0}(1+z)^{3}\;\nonumber \\
+(1-\Omega_{B}^{0}-\Omega_{DM}^{0}-\Omega_{r}^{0})(1+z)^{3(1+w_{0}+w_{1})}\;\nonumber \\
\times e^{-3w_{1}z/(1+z)}(1+\Theta(z,w_{0},w_{1},\delta))+\Omega_{r}^{0}(1+z)^{4}]^{1/2},\label{E(z)}
\end{eqnarray}
where $\Omega_{DM}^{0}$, $\Omega_{B}^{0}$ and
$\Omega_{r}^{0}$ are the dark matter, the baryonic and radiation
density fractions today, respectively. We adopt
$\Omega_{B}^{0}=0.0487 \pm 0.0006$
\citep{Planck2013arXiv1303.5076P}, $H_0 = 73.8 \pm 2.4 ~\textrm{km}
\textrm{s}^{-1}\textrm{Mpc}^{-1}$ \citep{Riess2011} and
$\Omega_{r}^{0}=(\Omega_{DM}^{0}+\Omega_{B}^{0})/(1+z_{eq})$, where
$z_{eq}$ is the redshift when matter energy density is equal to
radiation energy density.

\section{Observational data}

In order to constrain the parameters tightly, we combine SNe
Ia sample, Hubble parameter data, BAO measurements and CMB
observation. Each one of these data can constrain cosmological
parameter compactly and consistently
\citep{Suzuki2012,Farooq2013,Hinshaw2012,Planck2013arXiv1303.5076P}.

\subsection{SNe Ia data}

SNe Ia data is the first evidence for the accelerating expansion of the
universe, and it can be taken as standard candles to measure the
luminosity distance. We use the latest Union 2.1 sample
\citep{Suzuki2012}, which contains 580 SNe Ia in the redshift range
$0.014<z<1.415$. With the measured luminosity distance $d_{L}$ in
units of megaparsecs, the predicted distance modulus can be given as
\begin{equation}
\mu=5\log(d_{L})+25,
\end{equation}
where the luminosity distance is expressed as
\begin{equation}
d_{L}(z,\Omega_{DM}^{0},w_{0},w_{1},\delta)=c\frac{(1+z)}{H_{0}}\int_{0}^{z}\frac{dz'}{E(z',\Omega_{DM}^{0},w_{0},w_{1},\delta)}.
\end{equation}
The likelihood functions can be determined from $\chi_{SNe}^{2}$
distribution \citep{Nesseris2005,Wang2012},
\begin{equation}
\chi^{2}_{SNe}=A-\frac{B^2}{C},
\end{equation}
where $A=\sum_i^{580}{(\mu^{obs}-\mu^{th})^2}/{\sigma^2_{\mu,i}}$,
$B=\sum_i^{580}{(\mu^{obs}-\mu^{\rm th})}/{\sigma^2_{\mu,i}}$,
$C=\sum_i^{580}{1}/{\sigma^2_{\mu,i}}$. $\mu^{obs}$ is the
observational distance modulus, and $\sigma_{\mu,i}$ is the $1
\sigma$ uncertainty of the distance moduli.

\subsection{Hubble parameter data}

The Hubble parameter sample contains 28 data points, which cover redshift range
$0.07 \leq z \leq 2.3$. This is the largest data set of H(z), with
nine data from \cite{Simon2005}, two from
\cite{Stern2010}, eight from \cite{Moresco2012},
one from \cite{Busca2013}, four from \cite{Zhang2012},
three from \cite{Blake2012}, and one from
\cite{Chuang2013}. These data have been compiled by
\cite{Farooq2013} (see their Table 1). The $\chi^2_H$ is
given as
\begin{equation}
\chi^2_{H}=\sum_{i=1}^{28}\frac{[H(z_i)-H_{obs}(z_i)]^2}{\sigma_{h,i}^2},
\end{equation}
where theoretical $H(z)$ can be obtained from Eq. (\ref{E(z)}),
$H_{obs}$ and $\sigma_{h,i}$ are observed value.

\subsection{Baryon Acoustic Oscillations}

The BAO peak in galaxy correlation function is first detected in the
2dFGRS \citep{Cole2005} and SDSS \citep{Eisenstein2005}. Now the
BAO redshift covers the range $0.1 \leq z \leq 0.73$. The distance
ratio $d_z$ is defined as
\begin{equation}
d_z=\frac{r_s(z_d)}{D_V(z_{\mathrm{BAO}})},
\end{equation}
where the angular diameter distance scale $D_V$ is given by \cite{Eisenstein2005},
\begin{equation} D_V(z_{\mathrm{BAO}})=\frac{1}{H_0}\big
[\frac{z_{\mathrm{BAO}}}{E(z_{\mathrm{BAO}})}\big(\int_0^{z_{\mathrm{BAO}}}\frac{dz}{E(z)}\big
)^2\big]^{1/3}~.
\end{equation}

The comoving sound horizon at the drag epoch is
$r_s(z_d)={H_0}^{-1}\int_{z_d}^{\infty}c_s(z)/E(z)dz$. Following
\cite{Eisenstein1998}, the decouple redshift is
\begin{eqnarray}
z_{d}&=\{{1291(\Omega_{M}^{0}h^2)^{0.251}}/{[1+0.659(\Omega_{M}^{0}h^2)^{0.828}]}\}\;\nonumber
\\
&\times[(1+b_{1}(\Omega_{B}^{0}h^2)^{b_2})],
\end{eqnarray}
with
\begin{eqnarray}
b_1=0.313(\Omega_{M}^{0}h^2)^{-0.419}[1+0.607(\Omega_{M}^{0}h^2)^{0.674}]^{-1},\\
b_2=0.238(\Omega_{M}^{0}h^2)^{0.223}.
\end{eqnarray}

Here we will use the results from four data sets: 6dF Galaxy
Redshift Survey measurements at efficient redshift $z_{eff} = 0.1$
\citep{Beutler2011}, the SDSS DR7 BAO measurements at $z_{eff} =
0.35$ \citep{Padmanabhan2012}, the BOSS DR 9 measurements at
$z_{eff} = 0.57$ \citep{Anderson2012}, and WiggleZ measurements at
higher redshift $z_{eff} = 0.44, 0.60, 0.73$ \citep{Blake2012}.

The distance ratio vector is
\begin{center}
\begin{eqnarray}
\hspace{-.5cm}
~~~~~~~~~~~~{\bf{P}}_{\rm{BAO}}^{obs} = \left(\begin{array}{c}
{ d_{0.1}} \\
{d_{0.35}^{-1}}\\
{d_{0.57}^{-1}}\\
{ d_{0.44}} \\
{ d_{0.60}} \\
{ d_{0.73}} \\
\end{array}
  \right)=
  \left(\begin{array}{c}
0.336 \\
8.88\\
13.67\\
0.0916\\
0.0726\\
0.0592\\
\end{array}
  \right).
 \end{eqnarray}
\end{center}
The corresponding inverse covariance matrix is
\begin{eqnarray}
\hspace{-.3cm} &{\bf C_{\mathrm{BAO}}}^{-1}=
&\left(\begin{array}{cc}
\textbf{$I_{1}$} & 0 \\
0 & \textbf{$I_{2}$}  \\
\end{array}
\right),
\end{eqnarray}
where
\begin{eqnarray}
\hspace{-.3cm} &{\bf I_{1}}=
&\left(\begin{array}{ccc}
4444.4 & 0 & 0  \\
0 & 34.602 & 0  \\
0 & 0 & 20.661157\\
\end{array}
\right),
\end{eqnarray}

\begin{eqnarray}
\hspace{-.3cm} &{\bf I_{2}}=
&\left(\begin{array}{ccc}
24532.1  & -25137.7 & 12099.1  \\
-25137.7 & 134598.4 & -64783.9  \\
12099.1 & -64783.9 & 128837.6\\
\end{array}
\right).
\end{eqnarray}

The $\chi^2_{\mathrm{BAO}}$ value of the BAO can express as
\begin{equation}
\chi^2_{\mathrm{BAO}}=\Delta \textbf{P}_{\mathrm{BAO}}^{\mathrm{T}} C_{\mathrm{BAO}}^{-1} \Delta \textbf{P}_{\mathrm{BAO}},
\end{equation}
where $\Delta \textbf{P}_{\mathrm{BAO}}=\textbf{P}_{\mathrm{BAO}}^{th}-\textbf{P}_{\mathrm{BAO}}^{obs}$.

\subsection{CMB from WMAP 9 years}
We also use the WMAP 9 years data. We use the `` WMAP distance
priors" likelihood of 3 variables: the acoustic scale $l_a$, the
shift parameter $R$, and the recombination redshift $z_{\ast}$ to
constrain parameters. They can be expressed as
\begin{eqnarray}
l_a=\pi\frac{\int_0^{z_{\ast}}\frac{dz}{E(z)}/H_0}{r_s(z_{\ast})},\\
R=\frac{\Omega_{\mathrm{M0}}^{1/2} H_0}{c} \int_0^{z_{\ast}}\frac{dz}{E(z)},
\end{eqnarray}
and the recombination redshift is are given by \cite{Hu1996},
\begin{equation}
z_{\ast}=1048[1+0.00124(\Omega_B^{0}h^2)^{-0.738}(1+g_{1}(\Omega_{M}^{0}h^2)^{g_2})],\\
\end{equation}
with
\begin{eqnarray}
g_1=0.0783(\Omega_B^{0}h^2)^{-0.238}(1+39.5(\Omega_B^{0}h^2)^{0.763})^{-1},\\
g_2=0.560(1+21.1(\Omega_B^{0}h^2)^{1.81})^{-1}.
\end{eqnarray}
The best fitted data are given by \cite{Hinshaw2012},
 \begin{eqnarray}
\hspace{-.5cm}
~~~~~~{\textbf{P}}_{\rm{CMB}}^{obs} &=& \left(\begin{array}{c}
{l_a} \\
{ R}\\
{ z_{\ast}}\end{array}
  \right)=
  \left(\begin{array}{c}
302.40 \\
1.7246 \\
1090.88  \end{array}
  \right).
 \end{eqnarray}
The corresponding inverse covariance matrix can be written as
\begin{eqnarray}
\hspace{-.5cm}
~~~~~~{ \textbf{C}_{\mathrm{CMB}}}^{-1}=\left(
\begin{array}{ccc}
3.182 &18.253 &-1.429\\
18.253 &11887.879 &-193.808\\
-1.429 &-193.808 &4.556
\end{array}
\right).
\end{eqnarray}
The $\chi^2_{\mathrm{CMB}}$ value of CMB is
\begin{eqnarray}
\chi^2_{\mathrm{CMB}}=\Delta \textbf{P}_{\mathrm{CMB}}^{\mathrm{T}} \textbf{C}_{\mathrm{CMB}}^{-1} \Delta \textbf{P}_{\mathrm{CMB}},
\end{eqnarray}
where $\Delta \textbf{P}_{\mathrm{CMB}} =\textbf{P}_{\mathrm{CMB}}^{th}-\textbf{P}_{\mathrm{CMB}}^{obs}$.

\section{Methods and Results}
With the joint data, the total $\chi^2$ can be expressed as
\begin{equation}
\chi^2 (\delta,\Omega_{DM},w_0,w_1)=\chi^2_{SNe}+\chi^2_{H}+\chi^2_{BAO}+\chi^2_{CMB}.\label{chi}
\end{equation}
The model parameters can be determined by computing the $\chi^2$
distribution. First, we calculate the minimum value of the
total $\chi^2/dof=0.990$ from simultaneous fitting. Then, we
calculate the inverse covariance matrix to give out the best-fitted
value's $1 \sigma$ uncertainty£¬ $\delta=-0.022 \pm 0.006$,
$\Omega_{DM}^{0}=0.213 \pm 0.008$, $w_0 =-1.210 \pm 0.033$ and
$w_1=0.872 \pm 0.072$.

\begin{figure}
\begin{center}
\includegraphics[width=0.5\textwidth]{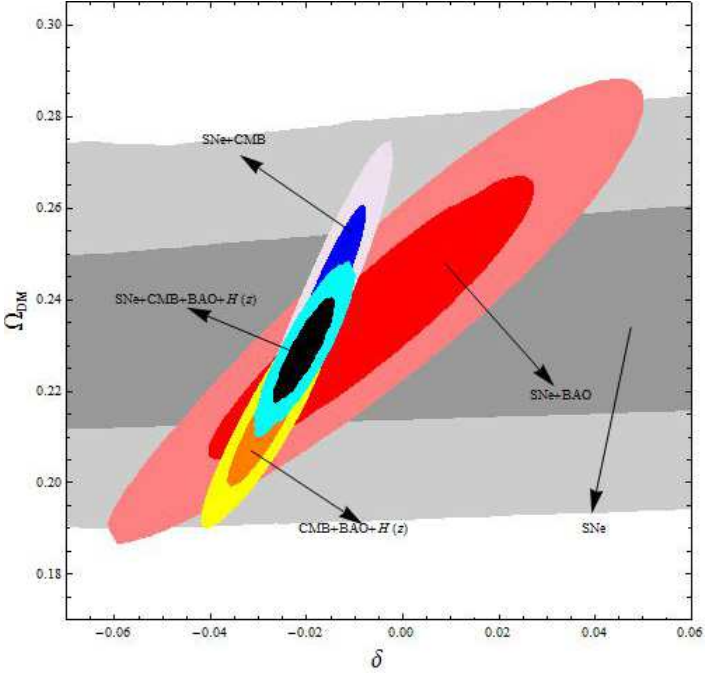}
\end{center}
\caption{The $\delta - \Omega_{DM}$ contours with different data combinations:
SNe (gray and light gray contours), SNe + BAO (red and pink contours),
SNe + CMB (blue and light purple contours), CMB + BAO + H(z) (Orange and
yellow contours) and SNe + CMB + BAO + H(z) (black and cyan contours). The
central regions and the vicinity regions represent $1 \sigma$ contours and $2 \sigma$
contours, respectively. \label{fig:contourdeom_fig}}
\end{figure}

In order to obtain the contour plot, we marginalize over other two
parameters to get a new $\chi^2$ function depending on two left
parameters,
\begin{equation}
\chi'^2 (\delta,\Omega_{DM})=\frac{1}{\Psi} \int_{w_0-\sigma_{w_0}}^{w_0+\sigma_{w_0}} \int_{w_1-\sigma_{w_1}}^{w_1+\sigma_{w_1}} \chi^2 (\delta,\Omega_{DM},w_0,w_1) d w_0 d w_1,\label{newchi}
\end{equation}
where $\Psi$ is the normalization factor to make the $\chi'^2$ have
the same minimum value as $\chi^2$. Then use $\chi'^2$ to give the
$\delta - \Omega_{DM}$ 2D marginalized regions with different colors
representing $1\sigma$ and $2\sigma$ regions. Figure
\ref{fig:contourdeom_fig} shows the $\delta -\Omega_{DM}$ contours
with different data combinations: SNe (gray and light gray
contours), SNe + BAO (red and pink contours), SNe + CMB (blue and
light purple contours), CMB + BAO + H(z) (Orange and yellow
contours) and the full data sets (black and cyan contours). This
figure shows that the BAO data can set tight constraint on
$\Omega_{DM}$, while CMB can set tighter constraint on $\delta$ and
$\Omega_{DM}$. From Figure 1, we find that there is a tension
between the SNe data and other data sets. The tension has been
investigated by \cite{Nesseris2005} and \cite{Wei2010 0}.

\begin{figure}
\begin{center}
\includegraphics[width=0.5\textwidth]{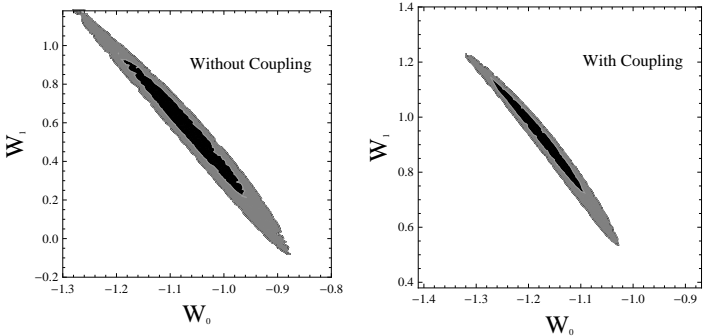}
\end{center}
\caption{The black and grey regions are $1 \sigma$ contours and $2
\sigma$ contours, respectively. The left panel is $w_0$ vs $w_1$ without coupling,
 and the right panel is $w_0$ vs $w_1$ with coupling in our model.
\label{fig:contour1_fig}}
\end{figure}

\begin{figure}
\begin{center}
\includegraphics[width=0.5\textwidth]{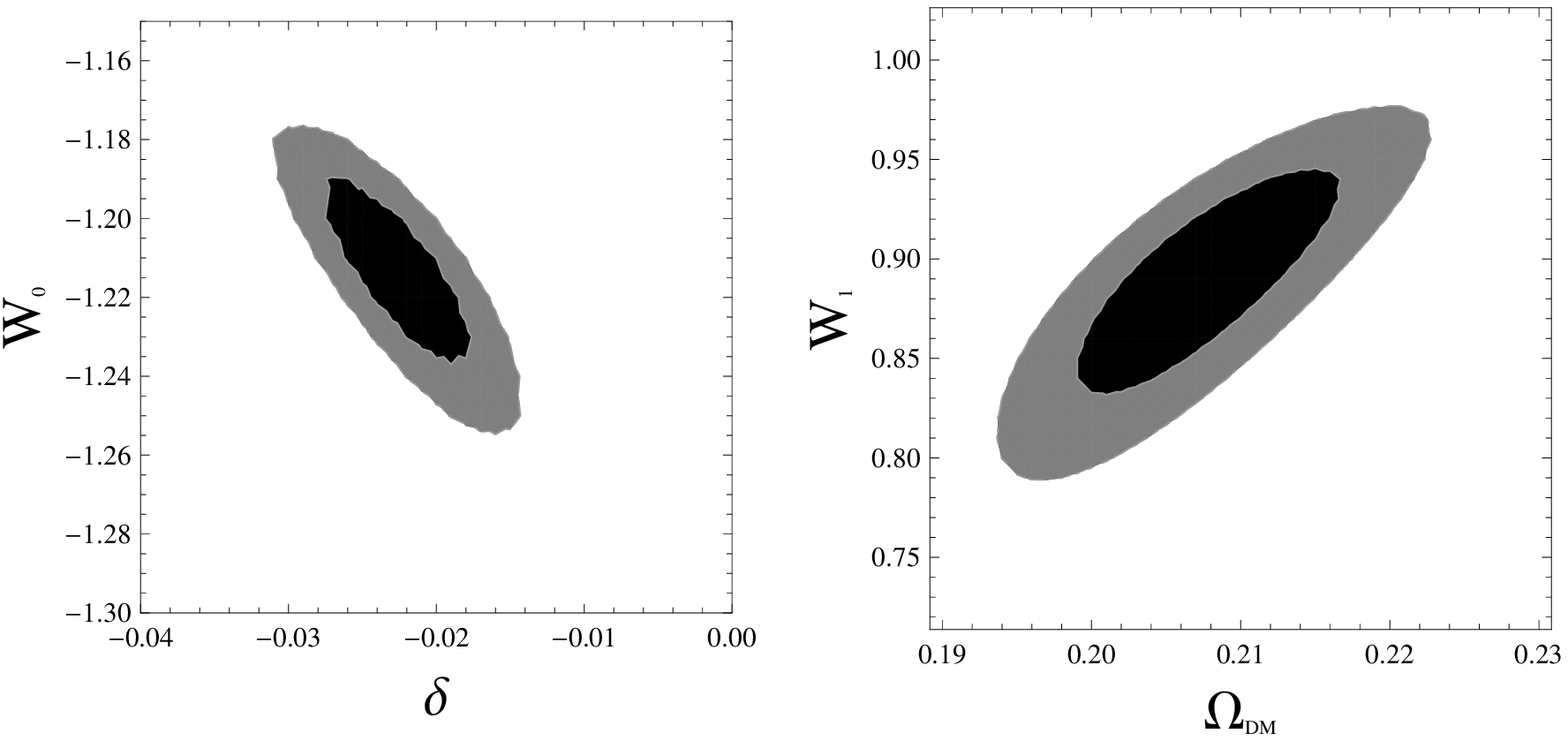}
\end{center}
\caption{The black and grey regions are $1 \sigma$ contours and $2
\sigma$ contours, respectively. The left panel is $\delta$ vs $w_0$,
and the right panel is $\Omega_{DM}$ vs $w_1$.
\label{fig:contour2_fig}}
\end{figure}

\begin{figure}
\begin{center}
\includegraphics[width=0.5\textwidth]{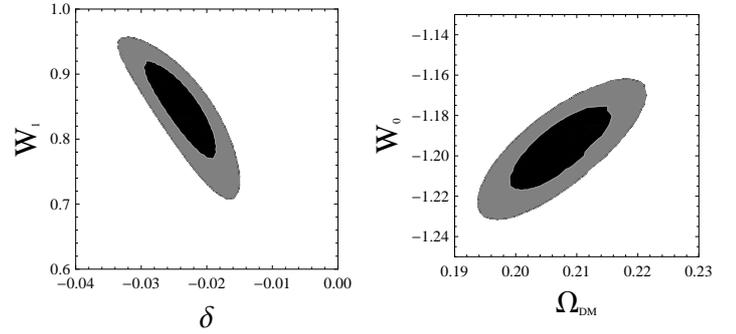}
\end{center}
\caption{The black and grey regions are $1 \sigma$ contours and $2
\sigma$ contours, respectively. The left panel is $\delta$ vs $w_1$,
and the right panel is $\Omega_{DM}$ vs $w_0$.
\label{fig:contour3_fig}}
\end{figure}

In order to test the reliability of our method, we also show
$w_0 - w_1$ contours from SNe+BAO+CMB without considering coupling
($\delta=0$) with $1\sigma$ in black region and $2\sigma$ in grey region
contours, which is presented in the left panel of Figure
\ref{fig:contour1_fig}. We can see that the our result is consistent
with that of WMAP team by comparing this figure with the Figure 10
of \cite{Hinshaw2012}. The right panel of Figure
\ref{fig:contour1_fig} shows the $w_0 - w_1$ contours with coupling
. We show $\delta - w_0$ and $\Omega_{DM} - w_1$ contours in Figure
\ref{fig:contour2_fig} and $\delta - w_1$ and $\Omega_{DM} - w_0$
contours in Figure \ref{fig:contour3_fig}, respectively.

\begin{figure}
\begin{center}
\includegraphics[width=0.5\textwidth]{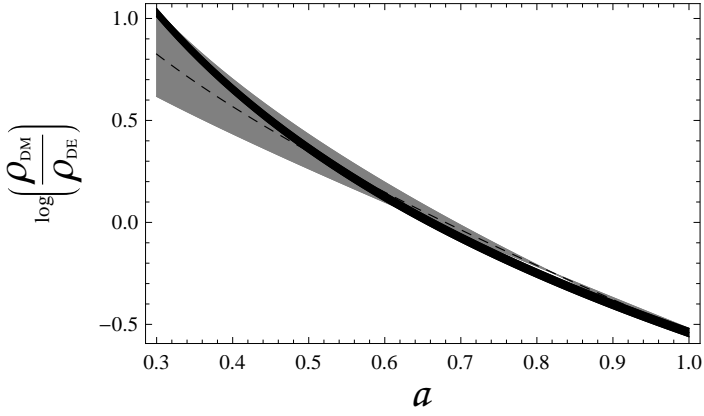}
\end{center}
\caption{The evolution of $\rho_{DM}/\rho_{DE}$ as a function of scale factor $a(z)$.
The dashed line is the interacting model with best-fitted parameters, and the gray
region is the 1$\sigma$ uncertainties.
The black region represents the $\Lambda$CDM with uncertainties.
\label{fig:proportion_fig}}
\end{figure}

From the best-fitted parameters, the energy density evolution of DM
and DE can be calculated. The ratio of DM density and DE density is
\begin{equation}
\rho_{DM}/\rho_{DE}=\rho_{DM}^{0}a^{-3+\delta}/(\rho_{DE}^{NI}(z)\left[1+\Theta(z,w_{0},w_{1},\delta)\right]).\label{DM/DE}
\end{equation}
Figure \ref{fig:proportion_fig} shows the evolution of $\rho_{DM} /
\rho_{DE}$ as a function of scale factor $a$ with best-fitted
parameters. The gray region is the $1 \sigma$ uncertainty in this
model, and the black one is for the $\Lambda$CDM case. In our model,
$\delta<0$ means that the energy transfers from dark matter to dark
energy, which is consisted with \cite{Dalal2001} and \cite{Guo2007}.
Nevertheless, the energy density proportion evolves slower than that
in $\Lambda$CDM case within $1 \sigma$ uncertainties when $a<0.5$,
which means that our model can help to relieve the coincidence
problem significantly.

\begin{figure}
\begin{center}
\includegraphics[width=0.5\textwidth]{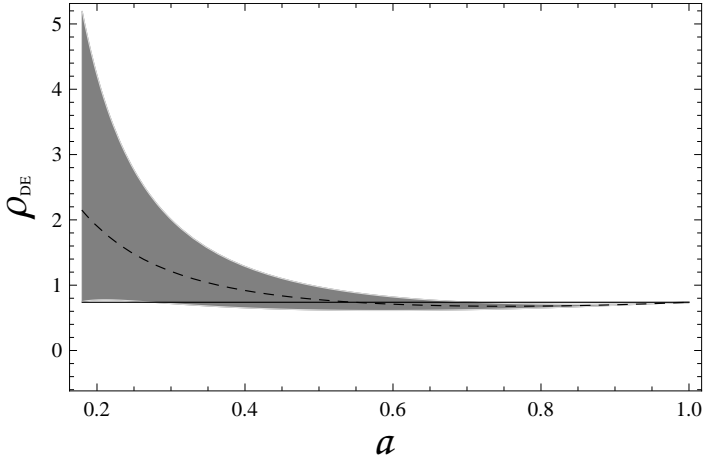}
\end{center}
\caption{The evolution of $\rho_{DE}$ as a function of scale factor $a(z)$.
The dashed line is the interacting model with best-fitted parameters, and the gray
region is the 1$\sigma$ uncertainties.
The black line represent the $\Lambda$CDM case.
\label{fig:deevofig}}
\end{figure}

The evolution of DE density plays an important role in
solving the coincidence problem. Using Eq.(\ref{DESol}) we can
compute the DE evolution, which is shown in Figure
\ref{fig:deevofig}. The gray region aloft the black line shows that
the DE density is decreasing within $1\sigma$ when $a<0.5$, which
can make the evolution of $\rho_{DM} / \rho_{DE}$ slower, resulting
in a good solution to the coincidence problem. However, the DE
density evolves quite quickly in the very early stage of the
universe when $a<0.3$. The main reason is that DM mass transfer rate
$\delta$ is assumed as a constant in our model.

Our universe is undergoing an accelerating expansion now. But in the
very early time, the universe was decelerating. So the evolution of
deceleration parameter $q(z)$ is important, especially when
$q(z_{tr})=0$, $z_{tr}$ is the transition redshift. $q(z)$ can be
expressed as
\begin{equation}
q=-\frac{a\ddot{a}}{\dot{a}^2} =-1+ \frac{1+z}{2 H(z)^2} \frac{d
H(z)^2}{dz}. \label{eq:deceleration parameter}
\end{equation}
After substituting the best-fitted parameters and their
uncertainties in Eq. (\ref{eq:deceleration parameter}), we obtain
$z_{tr}=0.63 \pm 0.07$. This value is a little bigger than those of
\cite{Wang2006}, \cite{Wang2007} and \cite{Abdel-Rahman2007} in
$\Lambda$CDM. The reason is that there exists an energy transfer
from DM to DE in our model. The DM density decreases quicker than
that in the $\Lambda$CDM, while the DE density is decreasing much
quicker in early times. So a higher transition redshfit is needed
for DE to oppose gravitation.

\section{Conclusions and Discussions}
In this paper, we use the Union 2.1 SNe Ia, CMB from WMAP 9 years,
BAO observation data from 6dFGRS, SDSS DR7, BOSS DR9, WiggleZ and
the latest Hubble parameter data to test the phenomenological
interacting dark sector scenario with a dynamic equation of state
$w_{DE}(z)=w_{0}+w_{1}z/(1+z)$. We give more stringent constraints
on the phenomenological model parameters: $\delta=-0.022 \pm 0.006$,
$\Omega_{DM}^{0}=0.213 \pm 0.008$, $w_0 =-1.210 \pm 0.033$ and
$w_1=0.872 \pm 0.072$ with $\chi^2_{min}/dof = 0.990$. From the
contours using different data combination in Figure 1, we find that
the SNe Ia are in tension with the CMB, BAO and Hubble parameter
data.

Our phenomenological scenario gives $\delta<0$ at $1\sigma$
confidence level, which is consistent with \cite{Dalal2001}
and \cite{Guo2007}. It indicates that the energy transfers
from dark matter to dark energy. But the evolution of
$\rho_{DM}/\rho_{DE}$ is slower than that in $\Lambda$CDM within
$1\sigma$ uncertainties, due to the $\rho_{DE}$ decreases with scale
factor $a$. So our model gives out a good approach to solve the
coincidence problem.

The DE density evolves quickly in very early epoch of the universe,
which is shown in Figure \ref{fig:deevofig}. The main reason is that
the value of $\delta$ is assumed to be constant in our model. In
real case, the DM mass transfer rate $\delta(a)$ should be varied.
We also derive the transition redshit $z_{tr}=0.63 \pm
0.07$ in this model. Due to the interaction between DE and DM, the
DE density decreases very quick in early times, so a higher
transition redshift is needed to resist gravitation.

\section*{Acknowledgments}
We thank the anonymous referee for helpful comments and suggestions
that have helped us improve our manuscript. This work is supported
by the National Basic Research Program of China (973 Program, grant
2014CB845800) and the National Natural Science Foundation of China
(grants 11373022, 11103007, 11033002 and J1210039).

\end{document}